\documentclass[twocolumn,final]{svjour3}

\usepackage{graphicx}  
\usepackage[round,authoryear]{natbib}
\usepackage{amsmath}  
\usepackage{hyperref}  
\usepackage[top=1.5in, bottom=1.5in, left=1in, right=1in]{geometry}
\usepackage{caption}
\usepackage{amsfonts}
\usepackage{algpseudocode}
\usepackage{algorithm}
\usepackage{color}

\newcommand{\w}{{\bf w}}

\title{Bayesian nonparametric spectral density estimation using B-spline priors}
\author{Matthew C. Edwards \and Renate Meyer \and Nelson Christensen}

\institute{Matthew C. Edwards \at Department of Statistics, University of Auckland, Auckland, New Zealand \\
Physics and Astronomy, Carleton College, Northfield, Minnesota, USA \\\email{matt.edwards@auckland.ac.nz} 
\and Renate Meyer  \at Department of Statistics, University of Auckland, Auckland, New Zealand 
\and Nelson Christensen \at Physics and Astronomy, Carleton College, Northfield, Minnesota, USA \\
Universit\'{e} C\^{o}te d'Azur, Observatoire de C\^{o}te d'Azur, CNRS, Artemis, Nice, France
}

\begin{document}\sloppy  

\maketitle

\begin{abstract}

    We present a new Bayesian nonparametric approach to estimating the
    spectral density of a stationary time series. A nonparametric
    prior based on a mixture of B-spline distributions is specified
    and can be regarded as a generalization of the Bernstein
    polynomial prior of \citet{petrone:1999a, petrone:1999b} and
    \citet{choudhuri:2004}.  Whittle's likelihood approximation is
    used to obtain the pseudo-posterior distribution.  This method
    allows for a data-driven choice of the number of mixture
    components and the location of knots.  Posterior samples are
    obtained using a Metropolis-within-Gibbs Markov chain Monte Carlo
    algorithm, and mixing is improved using parallel tempering.  We
    conduct a simulation study to demonstrate that for complicated
    spectral densities, the B-spline prior provides more accurate
    Monte Carlo estimates in terms of $L_1$-error and uniform coverage
    probabilities than the Bernstein polynomial prior.  We apply the
    algorithm to annual mean sunspot data to estimate the solar cycle.
    Finally, we demonstrate the algorithm's ability to estimate a
    spectral density with sharp features, using real gravitational
    wave detector data from LIGO's sixth science run, recoloured to
    match the Advanced LIGO target sensitivity.

\end{abstract}


\section{Introduction}\label{sec:intro}

Useful information about a stationary time series is encoded in its
spectral density, sometimes called the power spectral density (PSD).
This quantity describes the variance (or power) each individual
frequency component contributes to the overall variance of a time
series, and forms a Fourier transform pair with the autocovariance
function.  More formally, assuming an absolutely summable
autocovariance function ($\sum_{h = -\infty}^{\infty}|\gamma(h)| <
\infty$), the spectral density function $f(.)$ of a zero-mean weakly
stationary time series exists, is continuous and bounded, and is
defined as:
\begin{equation}
  \label{eq:spectral}
  f(\lambda) = \frac{1}{2\pi} \sum_{h = -\infty}^{\infty} \gamma(h)\exp(-\mathrm{i}h\lambda), \quad \lambda \in (-\pi, \pi],
\end{equation}
where $\lambda$ is angular frequency.  


Spectral density estimation methods can be broadly classified into two
groups: parametric and nonparametric.  Parametric approaches to
spectral density estimation are primarily based on autoregressive
moving average (ARMA) models \citep{brockwell:1991, barnett:1996},
but they tend to give misleading inferences when the parametric model
is poorly specified.

A large number of nonparametric estimation techniques are based on
smoothing the \textit{periodogram}, a process that randomly fluctuates
around the true PSD.  The periodogram, $I_n(.)$, is easily and
efficiently computed as the (normalized) squared modulus of Fourier
coefficients using the fast Fourier transform.  That is,

\begin{equation}
  \label{eq:periodogram}
  I_n(\lambda) = \frac{1}{2\pi n} \left| \sum_{t = 1}^n Y_t \exp(-\mathrm{i}t\lambda)\right|^2, \quad \lambda \in (-\pi, \pi], 
\end{equation}
where $\lambda$ is angular frequency, and $Y_t$ is a stationary time
series with discrete time points, $t = 1, 2, \ldots, n$.


Though the periodogram is an asymptotically unbiased estimator of the
spectral density, it is not a consistent estimator
\citep{brockwell:1991}.  Smoothing techniques such as Bartlett's
method \citep{bartlett:1950}, Welch's method \citep{welch:1967},
and the multitaper method \citep{thomson:1982} aim to reduce the
variance of the periodogram by dividing a time series into
(potentially overlapping) segments, calculating the periodogram for
each segment, and averaging over all of these.  Unfortunately, these
techniques are sensitive to the choice of smoothing parameter (i.e.,
the number of segments), resulting in a variance/bias trade-off.
Reducing the length of each segment also leads to lower frequency
resolution.

Another common nonparametric approach to spectral estimation involves
the use of splines.  Smoothing spline techniques are not new to
spectral estimation (see e.g., \citet{cogburn:1974} for an early
reference).  \citet{wahba:1980} used splines to smooth the
log-periodogram, with an automatic data-driven smoothing parameter,
avoiding the difficult problem of having to choose this quantity.
\citet{kooperberg:1995} used maximum likelihood and polynomial splines
to approximate the log-spectral density function.

{\em Bayesian} nonparametric approaches to spectrum estimation have
gained momentum in recent times.  In the context of splines,
\citet{gangopadhyay:1999} used a fixed low-order piecewise polynomial
to estimate the log-spectral density of a stationary time series.
They implemented a reversible jump Markov chain Monte Carlo (RJMCMC)
algorithm \citep{green:1995}, placing priors on the number of knots
and their locations, with the goal of estimating spectral densities
with sharp features.  \citet{choudhuri:2004} placed a Bernstein
polynomial prior \citep{petrone:1999a, petrone:1999b} on the spectral
density.  The Bernstein polynomial prior is essentially a finite
mixture of beta densities with weights induced by a Dirichlet process.
The number of mixture components is a smoothing parameter, chosen to
have a discrete prior.  \citet{zheng:2010} generalized this and
constructed a multi-dimensional Bernstein polynomial prior to estimate
the spectral density function of a random field.  Also extending the
work of \citet{choudhuri:2004}, \citet{macaro:2010} used informative
priors to extract unobserved spectral components in a time series, and
\citet{macaro:2014} generalized this to multiple time series.

Other interesting Bayesian nonparametric approaches include
\citet{carter:1997} inducing a prior on the log-spectral density using
an integrated Wiener process, and \citet{tonellato:2007} placing a
Gaussian random field prior on the log-spectral density.
\citet{liseo:2001}, \citet{rousseau:2012}, and \citet{chopin:2013}
used Bayesian nonparametric methods to estimate spectral densities
from long memory time series, and \citet{rosen:2012} focused on
time-varying spectra in nonstationary time series.

The majority of the Bayesian nonparametric methods (for short memory
time series) mentioned here make use of Whittle's approximation to the
Gaussian likelihood, often called the Whittle likelihood
\citep{whittle:1957}.  The Whittle likelihood, $L_n(.)$, for a
mean-centered weakly stationary time series $Y_t$ of length $n$ and
spectral density $f(.)$ has the following formulation:
\begin{equation}
  \label{eq:whittle}
  L_n(\mathbf{y} | f) \propto \exp\left(-\sum_{l = 1}^{\lfloor \frac{n-1}{2} \rfloor} \left(\log f(\lambda_l) + \frac{I_n(\lambda_l)}{f(\lambda_l)} \right) \right),
\end{equation}
where $\lambda_l = 2\pi l/n$ are the positive Fourier frequencies,
$\lfloor (n-1)/2 \rfloor$ is the greatest integer value less than or
equal to $(n-1)/2$, and $I_n(.)$ is the periodogram defined in
Equation~(\ref{eq:periodogram}).

The motivation for the work presented in this paper is to apply it in
signal searches for gravitational waves (GWs) using data from Advanced
LIGO \citep{ligo:2015} and Advanced Virgo \citep{virgo:2015}.  These
interferometric GW detectors have time-varying spectra, and it will be
important in future signal searches to be able to estimate the
parameters describing the noise simultaneously with the parameters of
a detected gravitational wave signal.  In a previous study
\citep{edwards:2015}, we utilized the methodology of
\citet{choudhuri:2004} to estimate the spectral density of simulated
Advanced LIGO \citep{ligo:2015} noise, while simultaneously estimating
the parameters of a rotating stellar core collapse GW signal.  The
method, based on the Bernstein polynomial prior, worked extremely well
on simulated data, but we found that it was not well-equipped to
detect the sharp and abrupt changes in an otherwise smooth spectral
density present in real LIGO noise \citep{christensen:2010,
  littenberg:2015}.  Under default noninformative priors, the method
tended to over-smooth the spectral density.  As detailed in Section
\ref{sec:approx}, this unsatisfactory performance is only partly due
to the well-known slow convergence of order $O(r^{-1/2})$, where $r$
is the degree of the Bernstein polynomials
\citep{PerronMengersen:2001}, but mainly due to a lack of coverage of
the space of spectral distributions by Bernstein polynomials.  This
can be overcome by using B-splines with variable knots instead of
Bernstein polynomials, yielding a much improved approximation of order
of $O(k^{-1})$ in the number of knots $k$ and adequate coverage of the
space of spectral distributions.



The focus of this paper is to describe a new Bayesian nonparametric
approach to modelling the spectral density of a stationary time
series.  Similar to \citet{gangopadhyay:1999}, our goal is to estimate
spectral density functions with sharp peaks, but the method is not
limited to these special cases.  Here we present an alternative
nonparametric prior using a mixture of B-spline densities, which we
will call the \textit{B-spline prior}.

Following \citet{choudhuri:2004}, we induce the weights for each of
the B-spline mixture densities using a Dirichlet process prior.
Furthermore, in order to allow for flexible, data-driven knot
placements, a second (independent) Dirichlet process prior is put on
the knot differences which, in turn, determines the shape and location
of the B-spline densities, and hence the structure of the spectral
density.  \citet{crandell:2011} applied a similar approach in the
context of functional data analysis.

A noninformative prior on the number of knots allows for a data-driven
choice of the smoothing parameter.  The B-spline prior could naturally
be interpreted as a generalization of the Bernstein polynomial prior,
as Bernstein polynomials are indeed a special case of B-splines where
there are no internal knots.

B-splines have the useful property of local support, where they are
only non-zero between their end knots.  We will demonstrate that if
knots are sufficiently close together, then the property of local
support will allow us to model sharp and abrupt changes to a spectral
density.

Samples from the pseudo-posterior distribution are obtained by
updating the B-spline prior with the Whittle likelihood
\citep{whittle:1957}.  This is implemented as a
Metropolis-within-Gibbs Markov chain Monte Carlo (MCMC) sampler
\citep{metropolis:1953, hastings:1970, geman:1984, gelman:2013}.  To
improve mixing and convergence, we use a parallel tempering scheme
\citep{swendsen:1986, earl:2005}.

We will demonstrate that the B-spline prior is more flexible than the
Bernstein polynomial prior and can better approximate sharp peaks in a
spectral density.  We will show that for complicated PSDs with
noninformative priors, the B-spline prior gives sensible Monte Carlo
estimates and outperforms the Bernstein polynomial prior in terms of
integrated absolute error (IAE) and frequentist uniform coverage
probabilities.  Furthermore, the placement of these knots is based on
the nonparametric Dirichlet process prior, meaning trans-dimensional
methods such as RJMCMC \citep{green:1995} can be avoided.  This is
useful as RJMCMC is often fraught with implementation difficulties,
such as finding an efficient jump proposal when there are indeed no
natural choices for trans-dimensional jumps \citep{brooks:2003}.

The paper is organized as follows.  Section~\ref{sec:bsplineprior}
sets out the notation and defines B-splines and B-spline
densities. After briefly reviewing the Bernstein polynomial prior, we
explain the rationale for the B-spline prior, extending it to a prior
for the spectral density of a stationary time series.  We discuss the
MCMC implementation in Section~\ref{sec:mcmc}.
Section~\ref{sec:simulation} details the results of the simulation
study, and in Section~\ref{sec:astronomy}, we apply the method to two
different astronomy problems.  This includes the annual mean sunspot
data set to estimate the duration of the solar cycle, and real
gravitational wave detector data to estimate a PSD with sharp
features.  Concluding remarks are then given in
Section~\ref{sec:conclusions}.




\section{The B-spline prior}\label{sec:bsplineprior}

In this section, the B-spline prior for the spectral density of a
stationary time series will be defined. To this end, we first set the
notation and define B-splines and B-spline densities. We review the
Bernstein polynomial prior and extend this approach to the B-spline
prior with variable knots.

\subsection{B-splines and B-spline densities}\label{sec:bspline}

A spline function of order $r+1$ is a piecewise polynomial of degree
$\leq r$ with so-called {\em knots} where the piecewise polynomials
connect. A spline is continuous at the knots (or continuously
differentiable to a certain order depending on the multiplicity of the
knots). The number of internal knots must be $\geq r $. Any spline
function of order $r+1$ defined on a certain partition can be uniquely
represented as a linear combination of basis splines,
\textit{B-splines}, of the same order over the same partition
\citep{powell:1981, cai:2011}.  B-splines can be parametrized either
recursively \citep{deBoor:1993}, or by using divided differences and
truncated power functions \citep{powell:1981, cai:2011}.  We will
adopt the former convention.  Without loss of generality, assume the
global domain of interest is the unit interval $[0, 1]$.

For a set of $k$ B-splines of degree $\leq r$ for some integer $r\geq
0$, define a nondecreasing knot sequence
\begin{eqnarray*}
  \boldsymbol\xi=&\{&0=\xi_0 =\xi_1 = \ldots =
  \xi_{r}\leq\xi_{r+1}\leq \ldots\leq\\ & & \xi_{k-1}\leq 1=\xi_{k}
  =\xi_{k+1}= \ldots =\xi_{k + r}\}\end{eqnarray*} of $k + r + 1$
  knots, comprised of $k - r + 1$ internal knots and $2r$ external
  knots.  The external knots outside or on the boundary of $[0, 1]$
  (i.e., $\xi_{0} \leq \ldots \leq \xi_{r - 1} \leq \xi_r = 0$ and $1
  = \xi_{k} \leq \xi_{k + 1} \ldots \leq \xi_{k + r}$) are required
  for B-splines to constitute a basis of spline functions on $[0, 1]$.
  Here we assume that the external knots are all exactly on the
  boundary.  The knot sequence $\boldsymbol\xi$ yields a partition of
  the interval $[0, 1]$ into $k-r$ subsets.

For $j = \{1, 2, \ldots, k\}$, each individual B-spline of degree $r$,
$B_{j, r}(.;\boldsymbol\xi)$, depends on $\boldsymbol\xi$ only through
the $r + 2$ consecutive knots $ (\xi_{j - 1}, \ldots, \xi_{j + r})$.
The number of internal knots is equal to the degree of the B-spline
$B_{j,r}$ if there are no knot multiplicities.  There can be a maximum
of $r + 1$ coincident knots for (right) continuity.  These knots
determine the shape and location of each B-spline.



A B-spline with degree 0 is the following indicator function
\begin{equation}
  \label{eq:bspline0}
  B_{j, 0}(\omega;\boldsymbol\xi) = 
  \begin{cases}
    1, \quad \omega \in [\xi_{j-1}, \xi_{j}), \\
    0, \quad \mathrm{otherwise}.
  \end{cases}
\end{equation}
Note that if $\xi_{j-1} = \xi_j$, then $B_{j, 0} = 0$.

Higher degree B-splines can then be defined recursively using
\begin{align}
  \label{eq:bspline_recursive}
  B_{j, r}(\omega;\boldsymbol\xi) &= \upsilon_{j, r}B_{j, r - 1}(\omega;\boldsymbol\xi) \nonumber \\
  &\quad + (1 - \upsilon_{j+1, r}) B_{j + 1, r - 1}(\omega;\boldsymbol\xi),
\end{align}
where $r > 0$ is the degree and
\begin{equation}
  \label{eq:test}
  \upsilon_{j, r} = 
  \begin{cases}
    \frac{\omega - \xi_{j-1}}{\xi_{j + r -1} - \xi_{j-1}}, & \xi_{j-1} \neq \xi_{j+r-1}, \\
    0, & \mathrm{otherwise}.
  \end{cases}
\end{equation}


\textit{B-spline densities} are the usual B-spline basis functions,
normalized so they each integrate to 1 \citep{cai:2011}.  The
recursive B-spline parametrization used in this paper allows us to
easily analytically integrate each B-spline, which we then use as
normalization constant for the B-spline density defined as
\begin{equation}
  \label{eq:bsplineDensity}
   b_{j, r}(\omega;\boldsymbol\xi) = \frac{B_{j, r}(\omega;\boldsymbol\xi)}{\int_{\xi_{j-1}}^{\xi_{j+r}} B_{j, r}(\omega;\boldsymbol\xi) d\omega}.
 \end{equation}




\subsection{Bernstein polynomial prior and B-spline prior}\label{sec:approx}

The Bernstein polynomial prior of \citet{petrone:1999a, petrone:1999b}
and \citet{choudhuri:2004} is based on the Weierstrass approximation
theorem that states that any continuous function on $[0, 1]$ can be
uniformly approximated to any desired accuracy by Bernstein
polynomials.  Let $G$ denote a cumulative distribution function (cdf)
with continuous density $g(.)$ on $[0, 1]$, then the following mixture
\begin{eqnarray*}
  \hat{G}(\omega)&=&\sum_{j=1}^r G\left(\frac{j-1}{r},\frac{j}{r}\right] I\beta(\omega;j,r-j+1)\\
  &=&\sum_{j=1}^r w_{j,r}  I\beta(\omega;j,r-j+1)
\end{eqnarray*}
converges uniformly to $G(\omega)$, where $G(u,v]=G(v)-G(u)$ and
$I\beta(\omega;a,b)$ and $\beta(\omega;a,b)$ denote the cdf and
density of the beta distribution with parameters $a$ and $b$,
respectively.

Define ${\cal F}=\{ F: F \mbox{ is a cdf on } [0,1]\}$ and ${\cal
  F}_r=\{ F: F \mbox{ is a mixture of } I\beta(j, r-j+1) \mbox{
  distributions}, j=1,\ldots,r\}$.  Also define the loss function
by \[ l({\cal F},{\cal F}_r)=\sup_{G\in {\cal F}} \inf_{F \in {\cal
    F}_r } \rho(G,F),\] where $\rho(G,F)=\sup_{x\in [0,1]}
|G(x)-F(x)|$. As shown by \citet{PerronMengersen:2001}, the loss
associated with the approximation of ${\cal F}$ by the $r-1$
dimensional space ${\cal F}_r$ with respect to loss function $l(.)$
cannot be made arbitrarily small.  Thus the mixture of beta cdfs does
not provide an adequate coverage of the space of cdfs on $[0, 1]$.
However, \citet{PerronMengersen:2001} showed that if one replaces the
beta distributions by B-spline distributions of fixed order (shown for
order 2, i.e., triangular distributions) but with variable knots, the
loss can be made arbitrarily small by increasing the number of knots.
This is the rationale for using a mixture of B-spline distributions
with variable knots in the following specification of a sieve prior.

The B-spline prior has the following representation as a mixture of
B-spline densities:
\begin{equation}
  \label{eq:bsplinemix1}
  s_r(\omega; k, \w_k,\boldsymbol\xi) = \sum_{j = 1}^k w_{j,k} b_{j,r}(\omega; \boldsymbol\xi),
\end{equation}
where $k$ is the number of B-spline densities of fixed degree $\leq r$
in the mixture, $\w_k=(w_{1,k},\ldots,w_{k,k})$ is the weight vector,
and $\boldsymbol\xi$ is the knot sequence. Rather than putting a prior
on the $\w_k$'s whose dimension changes with $k$, we follow the
approach of \citet{choudhuri:2004} and assume that the weights are
induced by a cdf $G$ on $[0, 1]$.  Similarly, we assume that the $k-r$
internal knot differences
$\displaystyle\Delta_j=\xi_{j+r}-\xi_{j+r-1}=H\left(\frac{j-1}{k-r},\frac{j}{k-r}\right]$
  for $j=\{1,\ldots,k-r\}$ are induced by a cdf $H$ on $[0, 1]$. Or
  equivalently, $\xi_{j+r}=H(\frac{j}{k-r})$ for $j=\{1,\ldots,k-r\}$,
  yielding the B-spline prior parametrized in terms of $k, G,$ and
  $H$:
\begin{equation}
  \label{eq:bsplinemix2}
  s_r(\omega; k, G,H) = \sum_{j = 1}^k G\left(\frac{j-1}{k},\frac{j}{k}\right]\ b_{j,r}(\omega; H).
\end{equation}
Independent Dirichlet process priors are then placed on $G$ and $H$
and a discrete prior is placed on the number of mixture components
$k$.



\begin{figure}
\includegraphics[width=1\linewidth]{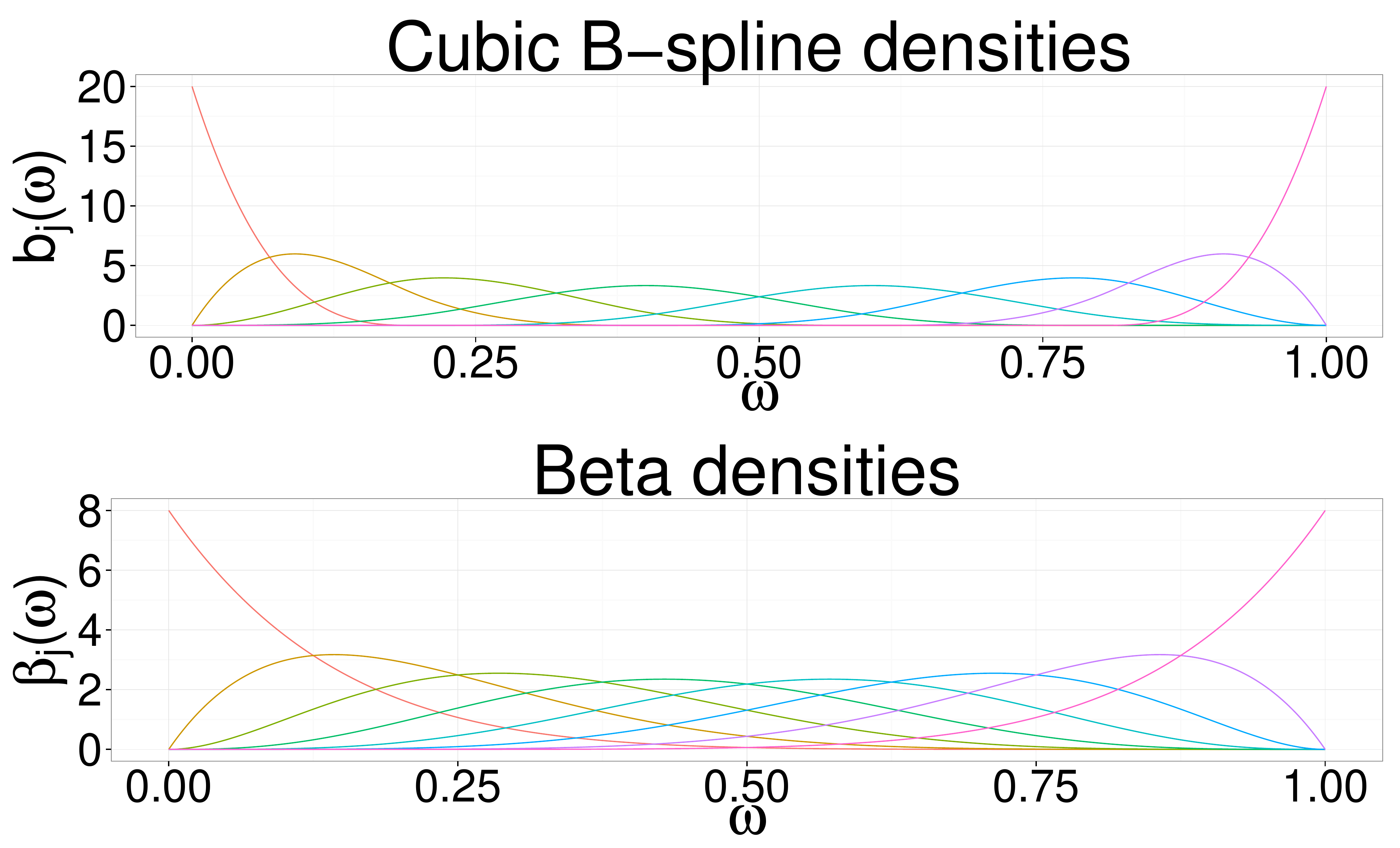}
\caption{Top panel: Eight cubic B-spline densities with equidistant
  knots at $\omega = \{0, 0.2, 0.4, 0.6, 0.8, 1\}$.  Notice the local
  support.  Bottom panel: Eight beta densities with full support on
  the entire unit interval.}
\label{fig:bsplineBernstein}
\end{figure}

The B-spline prior is similar in nature to the Bernstein polynomial
prior introduced by \citet{petrone:1999a, petrone:1999b} and applied
to spectral density estimation by \citet{choudhuri:2004}.  The primary
difference is that the B-spline prior is a mixture of B-spline
densities with \textit{local support} rather than beta densities with
\textit{full support} on the unit interval.  This difference is
illustrated in Figure~\ref{fig:bsplineBernstein}.

When there are no internal knots, the B-spline basis becomes a
Bernstein polynomial basis.  Bernstein polynomials are thus a special
case of B-splines, and the B-spline prior could be regarded as a
generalization of the Bernstein polynomial prior.

\begin{figure}
\includegraphics[width=1\linewidth]{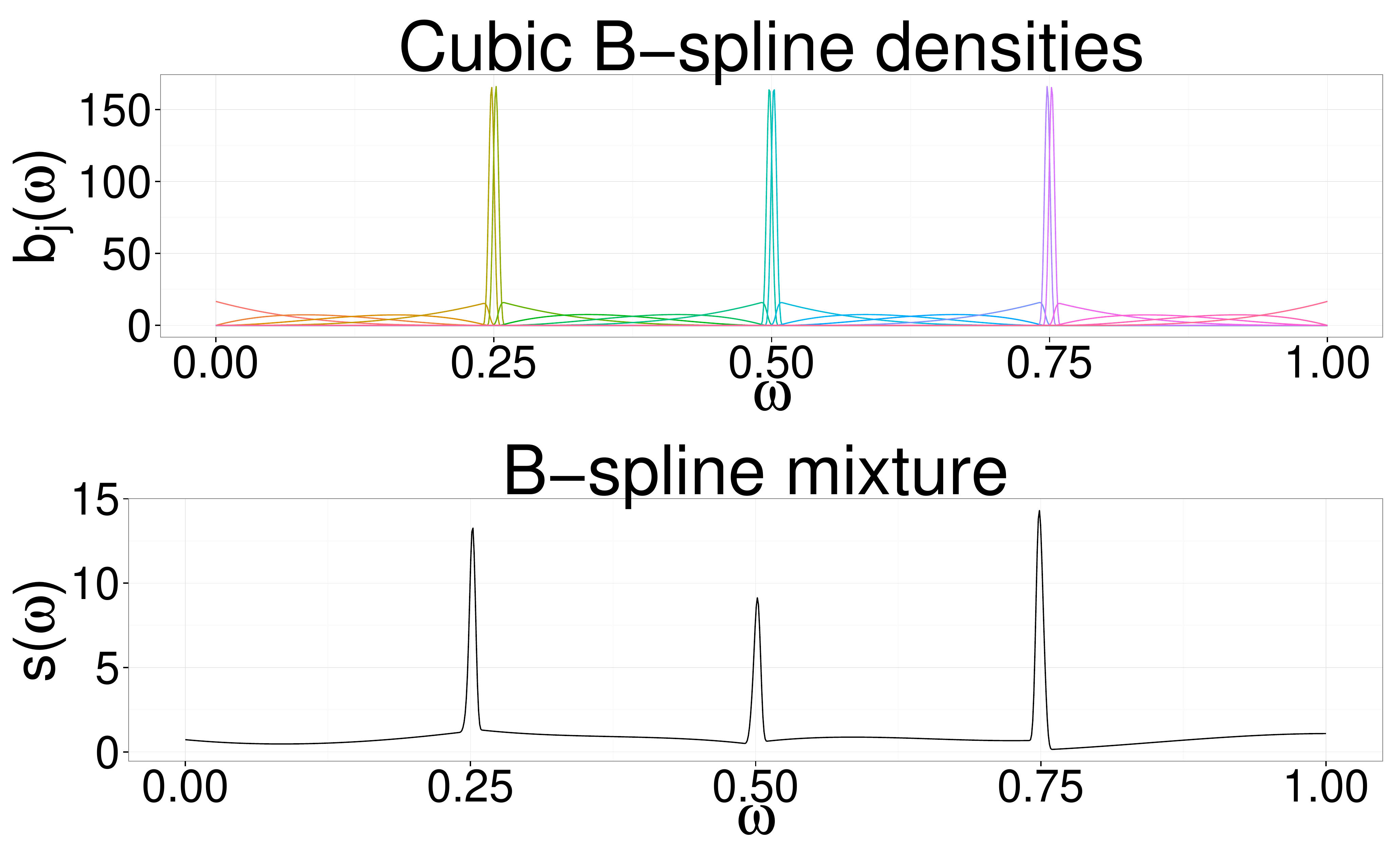}
\caption{Top panel: Cubic B-spline densities with many knots close to
  each of the locations $\omega = \{0.25, 0.5, 0.75\}$.  Bottom panel:
  A random mixture of these B-spline densities.  It is possible to
  construct a B-spline mixture with abrupt, sharp peaks.}
\label{fig:bsplineExample}
\end{figure}

Figure~\ref{fig:bsplineExample} demonstrates that it is possible to
construct curves (B-spline mixtures) with sharp peaks if knots are
sufficiently close together.  The top panel shows a set of B-spline
density functions and the bottom panel displays a mixture of these
with random weights.  The local support property of B-splines is the
reason the B-spline prior will be instrumental in estimating a
spectral density with sharp features.

\subsection{Prior for the spectral density}\label{sec:priorpsd}


To place a prior on the spectral density $f(.)$ of a stationary time
series defined on the interval $[0, \pi]$, we use the following
reparametrization:
\begin{equation}
  \label{eq:tauq}
  f(\pi\omega) = \tau \times s_r(\omega; k, G, H), \quad \omega \in [0, 1],
\end{equation}
where $\tau = \int_0^1 f(\pi\omega) \mathrm{d}\omega$ is the
normalization constant, and $s_r(.)$ is the B-spline prior
defined in Equation~(\ref{eq:bsplinemix2}).

The prior for $f(.)$ therefore has the following hierarchical
structure:
\begin{itemize}
\item $G$ determines the weights (i.e., scale) for each of the $k$
  B-spline densities.  Let $G \sim \mathrm{DP}(M_G, G_0)$, where $M_G
  > 0$ is the precision parameter and $G_0$ is the base probability
  distribution function with density $g_0$.
\item $H$ determines the location of knots and hence the shape and
  location of the B-spline densities.  Let $H \sim \mathrm{DP}(M_H,
  H_0)$, where $M_H > 0$ is the precision parameter and $H_0$ is the
  base probability distribution function with density $h_0$.
\item $k$ is the number of B-spline densities in the mixture (i.e.,
  smoothness) and has discrete probability mass function $p(k) \propto
  \mathrm{exp}(-\theta_k k^2)$ for $k = 1, 2, \ldots,
  k_{\mathrm{max}}$.  Here $k_{\mathrm{max}}$ is the largest possible
  value we allow $k$ to take.  We limit the maximum value of $k$ for
  computational reasons and do pilot runs to ensure a larger
  $k_{\max}$ is not required.  A smaller $k$ implies smoother spectral
  densities.
\item $\tau$ is the normalizing constant.  Let $\tau \sim
  \mathrm{IG}(\alpha_{\tau}, \beta_{\tau})$.
\end{itemize}

Assume all of these parameters are \textit{a priori} independent.


\section{Implementation using Markov chain Monte Carlo}\label{sec:mcmc}

As Dirichlet process priors have been placed on $G$ and $H$, we
require an algorithm to sample from these distributions.  To sample
from a Dirichlet process, we use Sethuraman's stick-breaking
construction \citep{sethuraman:1994}, an infinite-dimensional mixture
model.  For computational purposes, the number of mixture components
for the Dirichlet process representations of $G$ and $H$ is truncated
to large but finite positive integers ($L_G$ and $L_H$ respectively).
A larger choice of $L_G$ and $L_H$ will yield more accurate
approximations, but at the expense of increasing the computation time.

To set up the stick-breaking process, reparametrize $G$ to $(Z_0,
Z_1, \ldots, Z_{L_G}, V_1, \ldots, V_{L_G})$ such that
\begin{eqnarray}
  \label{eq:stickbreaking}
  G &=& \left(\sum_{l = 1}^{L_G} p_l\delta_{Z_l}\right) + \left(1 - \sum_{l = 1}^{L_G} p_l\right)\delta_{Z_0},\\
  p_1 &=& V_1, \\
  p_l &=& \left(\prod_{j = 1}^{l - 1}\left(1 - V_j\right)\right)V_l, \quad l \geq 2, \\
  p_0 &=& 1 - \sum_{l = 1}^{L_G} p_l, \\
  V_l &\sim& \mathrm{Beta}(1, M_G), \quad l = 1, \ldots, L_G, \\
  Z_l &\sim& G_0, \quad l = 0, 1, \ldots, L_G,
\end{eqnarray}\medskip

\noindent and $H$ to $(X_0, X_1, \ldots, X_{L_H}, U_1, \ldots, U_{L_H})$ such that
\begin{eqnarray}
  \label{eq:stickbreaking2}
  H &=& \left(\sum_{l = 1}^{L_H} q_l\delta_{X_l}\right) + \left(1 - \sum_{l = 1}^{L_H} q_l\right)\delta_{X_0}, \\
  q_1 &=& U_1, \\
  q_l &=& \left(\prod_{j = 1}^{l - 1}\left(1 - U_j\right)\right)U_l, \quad l \geq 2, \\
  q_0 &=& 1 - \sum_{l = 1}^{L_H} q_l, \\
  U_l &\sim& \mathrm{Beta}(1, M_H), \quad l = 1, \ldots, L_H, \\
  X_l &\sim& H_0, \quad l = 0, 1, \ldots, L_H, 
\end{eqnarray}
where $\delta_{a}$ is a probability density, degenerate at $a$,
i.e., $\delta_a = 1$ at $a$ and 0 otherwise.  

Conditional on $k$, the above hierarchical structure provides a finite
mixture prior for the spectral density of a stationary time series
\begin{equation}
  \label{eq:psdprior}
  f(\pi\omega) = \tau \sum_{j = 1}^k w_{j,k} b_{j,r}(\omega; \boldsymbol\xi),
\end{equation}
with weights 
\begin{equation}
  \label{eq:weights}
  w_{j,k} = \sum_{l = 0}^{L_G} p_l I\left\{\frac{j - 1}{k} <
Z_l \leq \frac{j}{k}\right\},
\end{equation}
and knot differences 
\begin{eqnarray}
  \label{eq:knotDiffs}
  \Delta_j &=& (\xi_{j + r} - \xi_{j + r - 1}) \\
&=& \sum_{l = 0}^{L_H} q_l I\left\{\frac{j - 1}{k - r} < X_l \leq \frac{j}{k - r}\right\},
\end{eqnarray}
for $j = \{1, \ldots, k - r\}$ and $k > r$.  The
denominator $k - r$ in the latter comes from assuming the exterior
knots are the same as the boundary knots.  
Note also that we assume the lower internal boundary knot $\xi_{r} =
0$, meaning the first knot difference is $\Delta_1 = \xi_{r + 1} -
\xi_{r} = \xi_{r + 1}$.  The subsequent knot placements are determined
by taking the cumulative sum of the knot differences.


Abbreviating the vector of parameters to $\boldsymbol\theta =
(\mathbf{v, z, u, x}, k, \tau)$, the joint prior is
\begin{align*}
  \label{eq:jointprior}
  p(\boldsymbol\theta) &\propto \left(\prod_{l = 1}^{L_G} M_G(1 - v_l)^{M_G-1}\right)\left(\prod_{l = 0}^{L_G} g_0(z_l)\right) \\
  &\quad \times \left(\prod_{l = 1}^{L_H} M_H(1 - u_l)^{M_H-1}\right)\left(\prod_{l = 0}^{L_H} h_0(x_l)\right) \\
  &\quad \times p(k)p(\tau).
\end{align*}

To produce the unnormalized joint pseudo-posterior, this joint prior
is updated using the Whittle likelihood defined in
Equation~(\ref{eq:whittle}).


We implement a Metropolis-within-Gibbs algorithm to sample points from
the pseudo-posterior, using the same modular symmetric proposal
distributions for B-spline weight parameters $\mathbf{V}$ and
$\mathbf{Z}$ as described by \citet{choudhuri:2004}.  That is, say for
$V_l$, propose a candidate from a uniform distribution with $[V_l -
\epsilon_l, V_l + \epsilon_l]$, modulo the circular unit interval.  If
the candidate is greater than 1, take the decimal part only, and if
the candidate is less than 0, add 1 to put it back into [0, 1].  This
is done for all of the $\mathbf{V}$ and $\mathbf{Z}$ parameters.
\citet{choudhuri:2004} found that $\epsilon_l = l / (l + 2\sqrt{n})$
worked well for most cases, and we also adopt this.  The same approach
is used analogously for the B-spline knot location parameters
$\mathbf{U}$ and $\mathbf{X}$.  Parameter $\tau$ has a conjugate
inverse-gamma prior and may be sampled directly.  Smoothing parameter
$k$ could be sampled directly from its discrete full conditional (as
done by \citet{choudhuri:2004}), though this can be computationally
expensive for large $k_{\max}$, so we use a Metropolis proposal
centered on the previous value of $k$, such that there is a 75\%
chance of jumping according to a discrete uniform on $[-1, 1]$, and a
25\% chance of boldly jumping according to a discretized Cauchy random
variable.

There is a common tendency towards multimodal posteriors in
finite/infinite mixture models.  If there are many isolated modes
separated by low posterior density, it is important to use a sampling
technique that mixes Markov chains efficiently, rather than relying on
the random walk behaviour of the Metropolis sampler.  In order to
mitigate poor mixing and to accelerate convergence of Markov chains,
we use parallel tempering, or replica exchange \citep{swendsen:1986,
  earl:2005} for the gravitational wave application in
Section~\ref{sec:ligo}.

The idea of parallel tempering is borrowed from physical chemistry,
where a system may be replicated multiple times at a series different
temperatures.  Higher temperature replicas are able to sample larger
volumes of the parameter space, whereas lower temperature replicas may
become stuck in local modes.  The method works by allowing the
exchange of information between neighbouring systems.  Information
from the high temperature replicas can trickle down to the low
temperature systems (including the posterior distribution of
interest), providing more representative posterior samples.

In the context of MCMC, parallel tempering involves introducing an
auxiliary variable called inverse-temperature, denoted $T_c^{-1}$ for
chains $c = \{1, 2, \ldots, C\}$.  This variable becomes an exponent
in the target distribution for each parallel chain, $p_c(.)$.  That
is, $p_c(\boldsymbol\theta | \mathbf{y}) ^ {T_c ^ {-1}}$, where
$\boldsymbol\theta$ are the model parameters, and $\mathbf{y}$ is the
time series data vector.  If $T_c^{-1} = 1$, this is the posterior
distribution of interest.  All other inverse-temperature values
produce tempered target distributions.  As $T_c \rightarrow \infty$,
the target distribution flattens out.  Each chain moves on its own in
parallel and occasionally swaps states between chains according to the
following Metropolis acceptance ratio:
\begin{equation}
  \label{eq:pt_accept}
  \varrho = \left(\frac{p(\boldsymbol\theta_j)p(\mathbf{y} | \boldsymbol\theta_j)}{p(\boldsymbol\theta_i)p(\mathbf{y} | \boldsymbol\theta_i)}\right)^{T_i^{-1} - T_j^{-1}},
\end{equation}
where information is exchanged between chains $i$ and $j$ and $i < j$.

We use cubic B-splines ($r = 3$) for all of the examples in the
following sections.  The serial version of the (cubic) B-spline prior
algorithm is available as a function called \textsf{gibbs\_bspline} in
the \textsf{R} package \textsf{bsplinePsd} \citep{bsplinePsd}.  This
is available on \textsf{CRAN}.  The parallel tempered version is
implemented in \textsf{R} using the \textsf{Rmpi} library but is not
publicly available.  Please contact the first author for access to
this code.



\section{Simulation study}\label{sec:simulation}

In this section, we run a simulation study on two autoregressive (AR)
time series of different order: AR(1) and AR(4).  For the first
scenario, an AR(1) with first order autocorrelation $\rho_1 = 0.9$
(a relatively simple spectral density) is generated.  In the second
scenario, an AR(4) with parameters $\rho_1~=~0.9, \rho_2~=~-0.9,
\rho_3~=~0.9$, and $\rho_4~=~-0.9$ is generated, such that the
spectral density has two large peaks.  Let each time series have
lengths $n~=~\{128,~256,~512\}$ with unit variance Gaussian
innovations.

We simulate 1,000 different realizations of AR(1) and AR(4) data and
model the spectral densities by running the Bernstein polynomial prior
algorithm of \citet{choudhuri:2004} and the B-spline prior algorithm
defined in Section~\ref{sec:mcmc} on each of these.  The MCMC
algorithms (without parallel tempering as mixing is satisfactory for
these toy examples) run for 400,000 iterations, with a burn-in period
of 200,000 and thinning factor of 10, resulting in 20,000 stored
samples.

For both spectral density estimation methods, we choose default
noninformative priors.  That is, for the B-spline prior, let $M_G =
M_H = 1, G_0 \sim \mathrm{Uniform}[0, 1], H_0 \sim \mathrm{Uniform}[0,
1], \theta_k = 0.01, \alpha_{\tau} = \beta_{\tau} = 0.001$.  For
comparability, we let the Bernstein polynomial prior have exactly the
same prior set-up as the B-spline prior, but obviously without knot
location parameter $M_H$ and distribution $H_0$.

We set $k_{\mathrm{max}} = 500$ for both algorithms.  This may seem
unnecessarily large for the B-spline prior algorithm as these simple
cases converge to a low $k$.  However, it is large enough to ensure
the Bernstein polynomial algorithm converges to an appropriate $k$,
without being truncated at $k_{\max}$.

Based on the suggestions by \citet{choudhuri:2004}, we set the
stick-breaking truncation parameters to $L_G = L_H = \max\{20,
n^{1/3}\}$.  This provides a reasonable balance between accuracy and
computational efficiency.

The (cubic) B-spline prior algorithm is run using the
\textsf{gibbs\_bspline} function in the \textsf{R} package
\textsf{bsplinePsd} \citep{bsplinePsd}.  The Bernstein polynomial
prior algorithm is run using the \textsf{gibbs\_NP} function in the
\textsf{R} package \textsf{beyondWhittle} \citep{kirch:2017,
  meier:2017}.  Both packages are available on \textsf{CRAN}.


An AR($p$) model has theoretical spectral density,
\begin{equation}
  \label{eq:arPSD}
  f(\lambda) = \frac{\sigma^2}{2\pi}\frac{1}{|1 - \sum_{j = 1}^p \rho_j \exp(-\mathrm{i}\lambda)|^2},
\end{equation}
where $\sigma^2$ is the variance of the white noise innovations and
$(\rho_1, \ldots, \rho_p)$ are the model parameters.  We can compare
estimates to this true spectral density to measure relative
performance of the nonparametric priors.  One measure of closeness and
accuracy is the integrated absolute error (IAE), also known as the
$L_1$-error.  This is defined as:
\begin{equation}
  \label{eq:iae}
  \mathrm{IAE} = \lVert \hat{f} - f \rVert_1 = \int_0^{\pi}|\hat{f}(\omega) - f(\omega)|d\omega,
\end{equation}
where $\hat{f}(.)$ is the Monte Carlo estimate (posterior median) of
the spectral density $f(.)$.  We calculate the IAE for each
replication and then compare the average IAE over all 1,000
replications.  The results are presented in Table~\ref{tab:iae}.


\begin{table}[!h]
    \begin{center}
    \caption{\label{tab:iae} Median $L_1$-error for the estimated
      spectral densities using B-spline prior and Bernstein
      polynomial prior on simulated AR(1) and AR(4) data.}
    \begin{tabular}{lccc}
    \hline
    AR(1)&$n = 128$&$n = 256$&$n = 512$\\
    \hline
    B-spline&0.901&0.756&0.592\\
    Bernstein&0.830&0.706&0.518\\
    \hline
    AR(4)&$n = 128$&$n = 256$&$n = 512$\\
    \hline
    B-spline&3.242&2.371&1.886\\
    Bernstein&3.427&2.920&2.656\\
    \hline
    \end{tabular}
  \end{center}
\end{table}

Table~\ref{tab:iae} compares the median IAE of the estimated spectral
densities under the two different nonparametric priors.  For the AR(1)
cases, the median IAE is only marginally higher for the B-spline prior
than the Bernstein polynomial prior.  As the AR(1) has a simple
spectral structure, this is a case where the global support of the
Bernstein polynomials makes sense.  However, when estimating the more
complicated AR(4) spectral density, we see that the B-spline prior
yields more accurate estimates than the Bernstein polynomial prior in
terms of IAE.  We also see that for both priors, as $n$ increases,
median IAE decreases.

For each simulation, we calculate two different credible regions: the
usual equal-tailed pointwise credible region, and the uniform (or
simultaneous) credible band \citep{neumann:1998a, neumann:1998b,
  lenhoff:1999, sun:1994}.  Uniform credible bands are very useful as
they allow the calculation of coverage levels for entire curves
(spectral densities in this case) rather than pointwise intervals.  To
compute a $100(1 - \alpha)$\% uniform credible band, we use the
following form:
\begin{equation}
  \label{eq:uniformCI}
  \hat{f}(\lambda) \pm \zeta_{\alpha} \times \mathrm{mad}(\hat{f_i}(\lambda)), \quad \lambda \in [0, \pi],
\end{equation}
where $\hat{f}(\lambda)$ is the pointwise posterior median spectral
density, $\mathrm{mad}(\hat{f_i}(\lambda))$ is the median absolute
deviation of the posterior samples $\hat{f_i}(\lambda)$ kept after
burn-in and thinning (which are used as the estimate of dispersion of
the sampling distribution of $\hat{f}(\lambda)$), and we choose the
$\zeta_{\alpha}$ such that
\begin{equation}
  \label{eq:uniformmax}
  \mathbb{P}\left\{\max\left\{\frac{|\hat{f_i}(\lambda) - \hat{f}(\lambda)|}{\mathrm{mad}(\hat{f_i}(\lambda))}\right\} \leq \zeta_{\alpha} \right\} = 1 - \alpha.
\end{equation}


Based on these uniform credible bands, uniform coverage probabilities
over all 1,000 replications of the simulation can be computed.  That
is, calculate the proportion of times that the true spectral density
is entirely encapsulated within the uniform credible band.  Computed
coverage probabilities are shown in Table~\ref{tab:ucp}.


\begin{table}[!h]
    \begin{center}
    \caption{\label{tab:ucp} Coverage probabilities based on 90\% uniform
      credible bands.}
    \begin{tabular}{lccc}
    \hline
    AR(1)&$n = 128$&$n = 256$&$n = 512$\\
    \hline
    B-spline&1.000&1.000&0.998\\
    Bernstein&1.000&0.987&0.499\\
    \hline
    AR(4)&$n = 128$&$n = 256$&$n = 512$\\
    \hline
    B-spline&0.936&0.979&0.907\\
    Bernstein&0.000&0.000&0.000\\
    \hline
    \end{tabular}
  \end{center}
\end{table}

\begin{figure}
\includegraphics[width=1\linewidth]{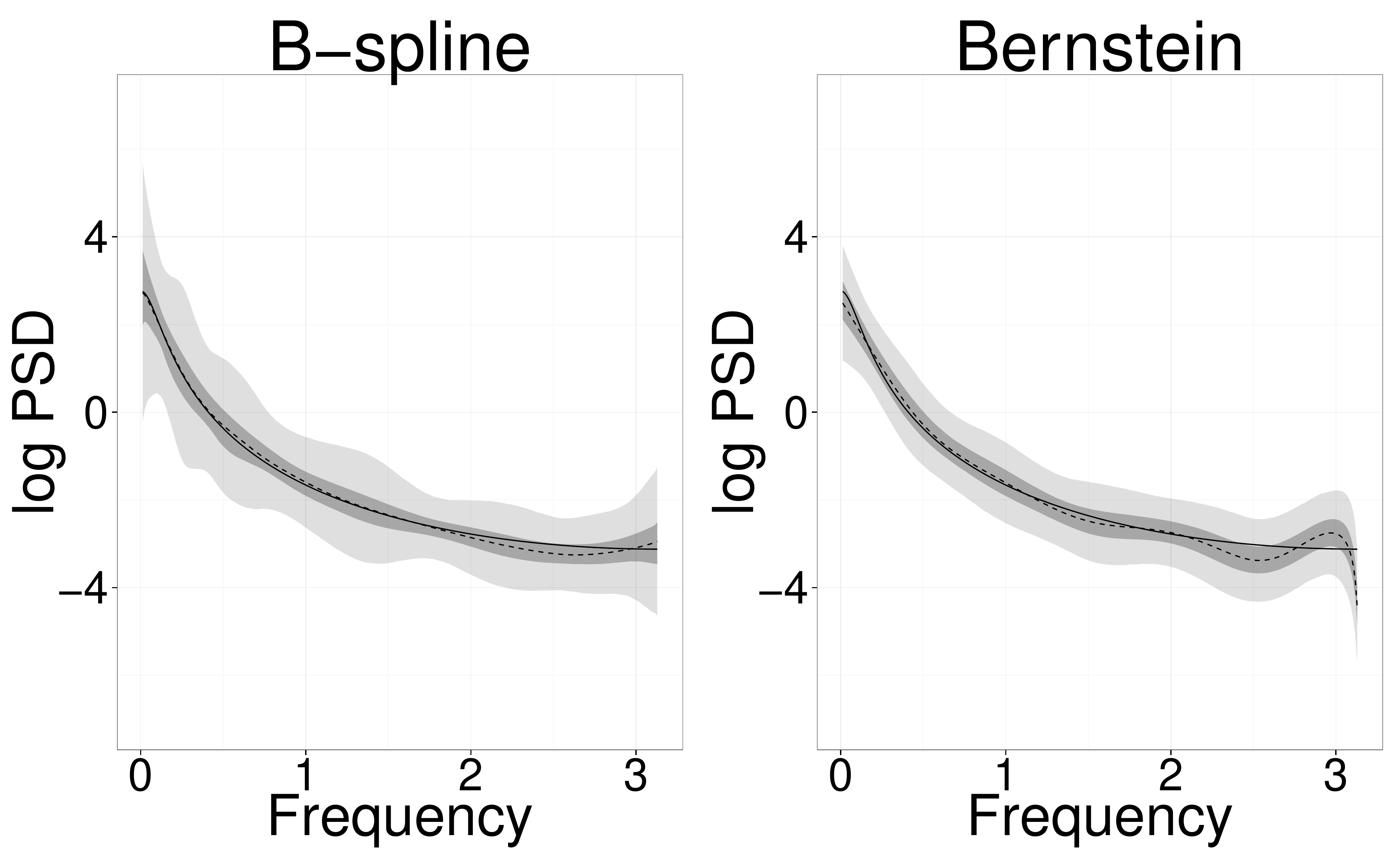}
\caption{Estimated log-spectral densities for an AR(1) time series
  using the B-spline prior (left) and Bernstein polynomial prior
  (right).  The solid line is the true log-PSD; the dashed line is the
  posterior median log-PSD; the dark shaded region is the pointwise
  90\% credible region; and the light shaded region is the uniform
  90\% credible band.}
\label{fig:ar1ci}
\end{figure}

It can be seen in Table~\ref{tab:ucp} that the B-spline prior has
higher coverage than the Bernstein polynomial prior in all examples
(apart from the AR(1) with $n = 128$, where it is the same).  The
B-spline prior produces excellent coverage probabilities for the AR(1)
cases.  The Bernstein polynomial prior also performs well in this
regard, apart from the $n = 512$ case, where half are not fully
covered.  An example from one of the 1,000 replications of the AR(1)
with $n = 512$ is given in Figure~\ref{fig:ar1ci}.  Here, the uniform
credible band fully contains the true PSD for the B-spline prior but
not for the Bernstein polynomial prior (the true PSD falls outside the
uniform credible band at the highest frequencies).  The pointwise
credible region and posterior median log-PSD for both priors are also
very accurate.  This is not surprising as the AR(1) has a relatively
simple spectral structure.

\begin{figure}
\includegraphics[width=1\linewidth]{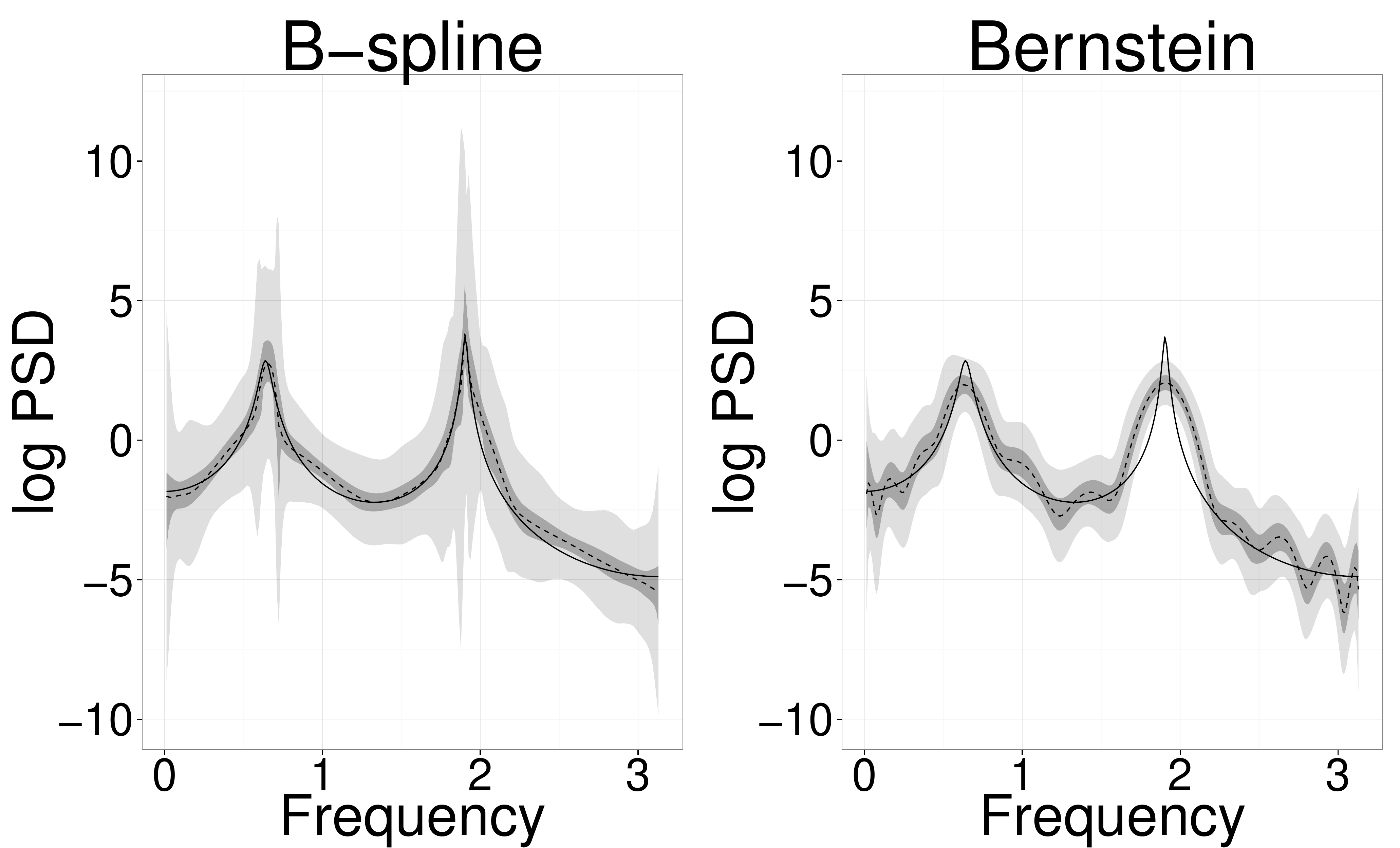}
\caption{Estimated log-spectral densities for an AR(4) time series
  using the B-spline prior (left) and Bernstein polynomial
  prior (right).  The solid line is the true log-PSD; the
  dashed line is the posterior median log-PSD; the dark shaded region
  is the pointwise 90\% credible region; and the light shaded region
  is the uniform 90\% credible band.}
\label{fig:ar4ci}
\end{figure}

Coverage of the AR(4) spectral density under the B-spline prior is
above 90\% for each sample size.  However, the Bernstein polynomial
prior has extremely poor coverage in the AR(4) case, where none of the
1,000 replications are fully covered by the uniform credible band for
each sample size.  An example of this performance (for $n = 512$) can
be seen in Figure~\ref{fig:ar4ci}.  The Bernstein polynomial prior
(under the noninformative prior set-up) tends to poorly estimate the
second large peak of the PSD, and introduces additional incorrect
peaks and troughs throughout the rest of estimate.  These false peaks
and troughs are present due to the Bernstein polynomial prior
algorithm converging to large $k$ in an attempt to approximate the two
large peaks of the AR(4) PSD.  The B-spline prior gives a much more
accurate Monte Carlo estimate.  The posterior median log-PSD is close
to the true AR(4) PSD, the 90\% pointwise credible region mostly
contains the true PSD, and the 90\% uniform credible band fully
contains it.

Of course, the Bernstein polynomial prior could perform better on
spectral densities with sharp features if significant prior knowledge
was known in advance.  This can however be a formidable task, and is
not very generalizable to other time series.  A benefit of using the
B-spline prior is its ability to estimate a variety of different
spectral densities using the default noninformative priors used in
this paper.  We will see more examples of this in
Section~\ref{sec:astronomy}.

One slight drawback of the B-spline prior algorithm is its
computational complexity relative to the Bernstein polynomial prior.
B-spline densities must be evaluated many times per MCMC iteration
(when sampling $k, \mathbf{U}$, and $\mathbf{X}$) due to the variable
knot placements, whereas beta densities can be pre-computed and stored
in memory, saving much computation time.

Table~\ref{tab:comptime} displays the median run-time (over each 1,000
replication) for each of the six AR simulations.
\begin{table}[!h]
    \begin{center}
      \caption{\label{tab:comptime} Median absolute run-times (hours)
        and their associated relative run-times.}
    \begin{tabular}{lccc}
    \hline
    AR(1)&$n = 128$&$n = 256$&$n = 512$\\
    \hline
    B-spline&2.967&3.186&3.659\\
    Bernstein&1.423&1.572&1.844\\
    B-spline/Bernstein&2.086&2.026&1.985\\
    \hline
    AR(4)&$n = 128$&$n = 256$&$n = 512$\\
    \hline
    B-spline&4.044&4.422&5.174\\
    Bernstein&1.443&1.694&2.281\\
    B-spline/Bernstein&2.802&2.610&2.268\\
    \hline
    \end{tabular}
  \end{center}
\end{table}

It can be seen in Table~\ref{tab:comptime} that the B-spline prior
algorithm is approximately 2--3 times slower than the Bernstein
polynomial prior algorithm for these examples.  Due to the noted
advantages that the B-spline prior has over the Bernstein polynomial
prior (such as accuracy and coverage), particularly for PSDs with
complicated structures, the increased computation time is an
acceptable trade-off, though for simple spectral densities, the
Bernstein polynomial prior should suffice.

\section{Applications in astronomy}\label{sec:astronomy}

\subsection{Annual sunspot numbers}\label{sec:sunspot}

In this section, we analyze the annual mean sunspot numbers from 1700
to 1987.  Sunspots are darker and cooler regions of the Sun's surface
caused by magnetic fields penetrating the surface from below.
Sunspots are linked to various solar phenomena such as solar flares
and the auroras.

Previous analyses have shown that the sunspot (or solar) cycle reaches
a solar maximum approximately every 11 years (see e.g.,
\citet{schwabe:1843} for the original reference and
\citet{choudhuri:2004} for analysis using the Bernstein polynomial
prior).  The analysis in this section is consistent with this claim.

As done by \citet{choudhuri:2004}, we first transform the data by
taking the square root of the original 288 observations to make the
data stationary.  We then mean-center the resulting data.

\begin{figure}
\includegraphics[width=1\linewidth]{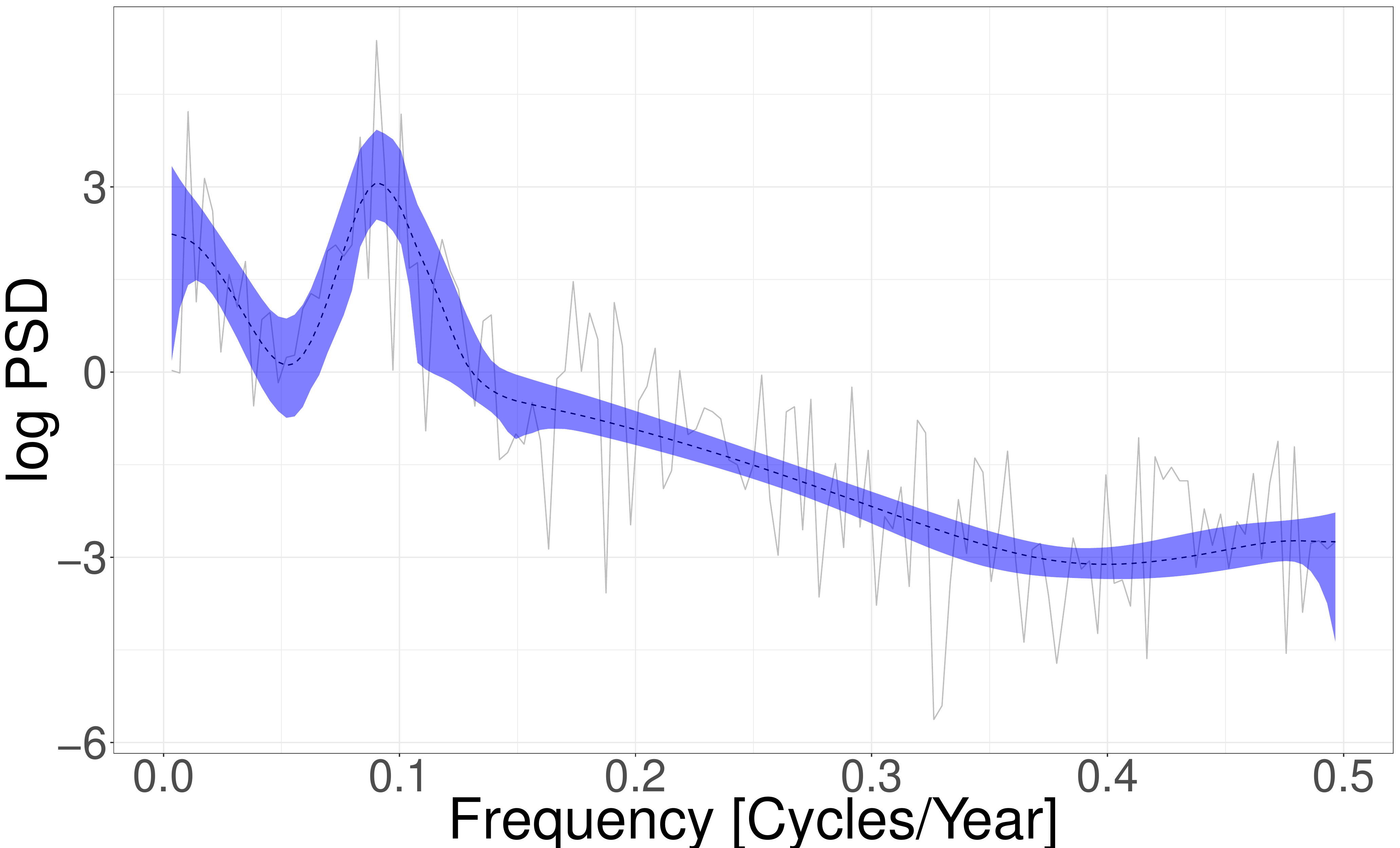}
\caption{Estimated log-PSD for the annual mean sunspot
  numbers from 1700 to 1987.  The posterior median log-PSD (dashed
  black) along with the 90\% pointwise credible region (shaded blue)
  are overlaid with the log-periodogram (grey).}
\label{fig:sunspot}
\end{figure}

The serial version of the B-spline prior MCMC algorithm is run for
100,000 iterations, with a burn-in period of 50,000 and thinning
factor of 10, resulting in 5,000 stored samples.  This takes
approximately 40 minutes to run.  All other specifications are the
same as in Section~\ref{sec:simulation}.  An estimate of the PSD is
displayed in Figure~\ref{fig:sunspot}.

It can be seen in Figure~\ref{fig:sunspot} that a spectral peak occurs
at the frequency of 0.0903 cycles per year.  This is equivalent to a
solar cycle every 11.07 years, consistent with existing knowledge.

\subsection{Recoloured LIGO gravitational wave data}\label{sec:ligo}

Gravitational waves (GWs) are ripples in the fabric of spacetime,
caused by the most exotic and cataclysmic astrophysical events in the
cosmos, such as binary black hole or neutron star mergers, core
collapse supernovae, pulsars, and even the Big Bang.  They are a
consequence of Einstein's general theory of relativity
\citep{einstein:1916}.


On September 14, 2015, the breakthrough first direct detection of GWs
was made using the Advanced LIGO detectors \citep{detection:2016}.
The signal, GW150914, came from a pair of stellar mass black holes
that coalesced approximately 1.3 billion light-years away.  This was
also the first direct observation of black hole mergers.  Four
subsequent detections of pairs of stellar mass black holes have been
made \citep{detection:2016b, detection:2017, detection:2017b,
  detection:2017c}, as well as the first binary neutron star detection
with an electromagnetic counterpart \citep{bns:2017}, signalling a new
era of astronomy is now upon us.

Advanced LIGO is a set of two GW interferometers in the United States
(one in Hanford, Washington, and the other in Livingston, Louisiana)
\citep{ligo:2015}.  Data collected by these observatories are
dominated by instrument and background noise --- primarily seismic,
thermal, and photon shot noise.  There are also high power, narrow
band, spectral noise lines caused by the AC electrical supplies and
mirror suspensions, among other phenomena.  Though GW150914 had a
large signal-to-noise ratio, signals detected by these observatories
will generally be relatively weak.  Improving the characterization of
detector/background noise could therefore positively impact signal
characterization and detection confidence.

The default noise model in the gravitational wave literature assumes
instrument noise is Gaussian, stationary, and has a known PSD.  Real
data often depart from these assumptions, motivating the development
of alternative statistical models for detector noise.  In the
literature, this includes Student-t likelihood generalizations by
\citet{roever:2011a} and \citet{roever:2011b}, introducing additional
scale parameters and marginalization by \citet{littenberg:2013} and
\citet{vitale:2014}, modelling the broadband PSD with a cubic spline
and spectral lines with Cauchy random variables by
\citet{littenberg:2015}, and the use of a Bernstein polynomial prior
by \citet{edwards:2015}.

We found that due to the undesirable properties of the Bernstein
polynomial prior, it was not flexible enough to estimate sharp peaks
in the spectral density of real LIGO noise.  This, coupled with the
fact that B-splines have local support, provided the rationale for
implementing the B-spline prior instead.

In the following example, using the parallel tempered B-spline prior
algorithm, we estimate the PSD of a 1~s stretch of real LIGO data
collected during the sixth science run (S6), recoloured to match the
target noise sensitivity of Advanced LIGO \citep{christensen:2010}.
LIGO has a sampling rate of 16384~Hz.  To reduce the volume of data
processed, a low-pass Butterworth filter (of order 10 and attenuation
0.25) is applied, then the data are downsampled to 4096~Hz (resulting
in a sample size of $n = 4096$).  Prior to downsampling, the data are
differenced once to become stationary, mean-centered, and then Hann
windowed to mitigate spectral leakage.  Though a 1~s stretch may seem
small in the context of GW data analysis, this time scale is important
for on-source characterization of noise during short-duration
transient events, called \textit{bursts} \citep{bursts:2012}.  This is
particularly true since LIGO noise has a time-varying spectrum, and
systematic biases could occur if off-source noise was used to estimate
the power spectrum of on-source noise.

We run 16 parallel chains (each at different temperatures) of the MCMC
algorithm for $400,000$ iterations, with a burn-in of $200,000$ and
thinning factor of 5, yielding $40,000$ stored samples.  We propose
swaps (of all parameters blocked together) between adjacent chains on
every tenth iteration.  For each chain $c$, we found the following
inverse-temperature scheme gave reasonable results:
\begin{equation}
  \label{eq:ladder}
  T_c ^ {-1} = T_{\mathrm{min}}^{-\Delta_c},
\end{equation}
where $T_{\mathrm{min}} = 0.005$ is the minimum inverse-temperature
allowed, $\Delta_c = \frac{c - 1}{C - 1}$, and $C = 16$ is the number
of chains.  The stick-breaking truncation parameters are set to $L_G =
L_H = 20$ and all of the other prior specifications are exactly the
same as used in the AR simulation study of
Section~\ref{sec:simulation}.  Note that as the sample size for this
example is very large ($n = 4096$), the algorithm took several hours
to run.

\begin{figure}
\includegraphics[width=1\linewidth]{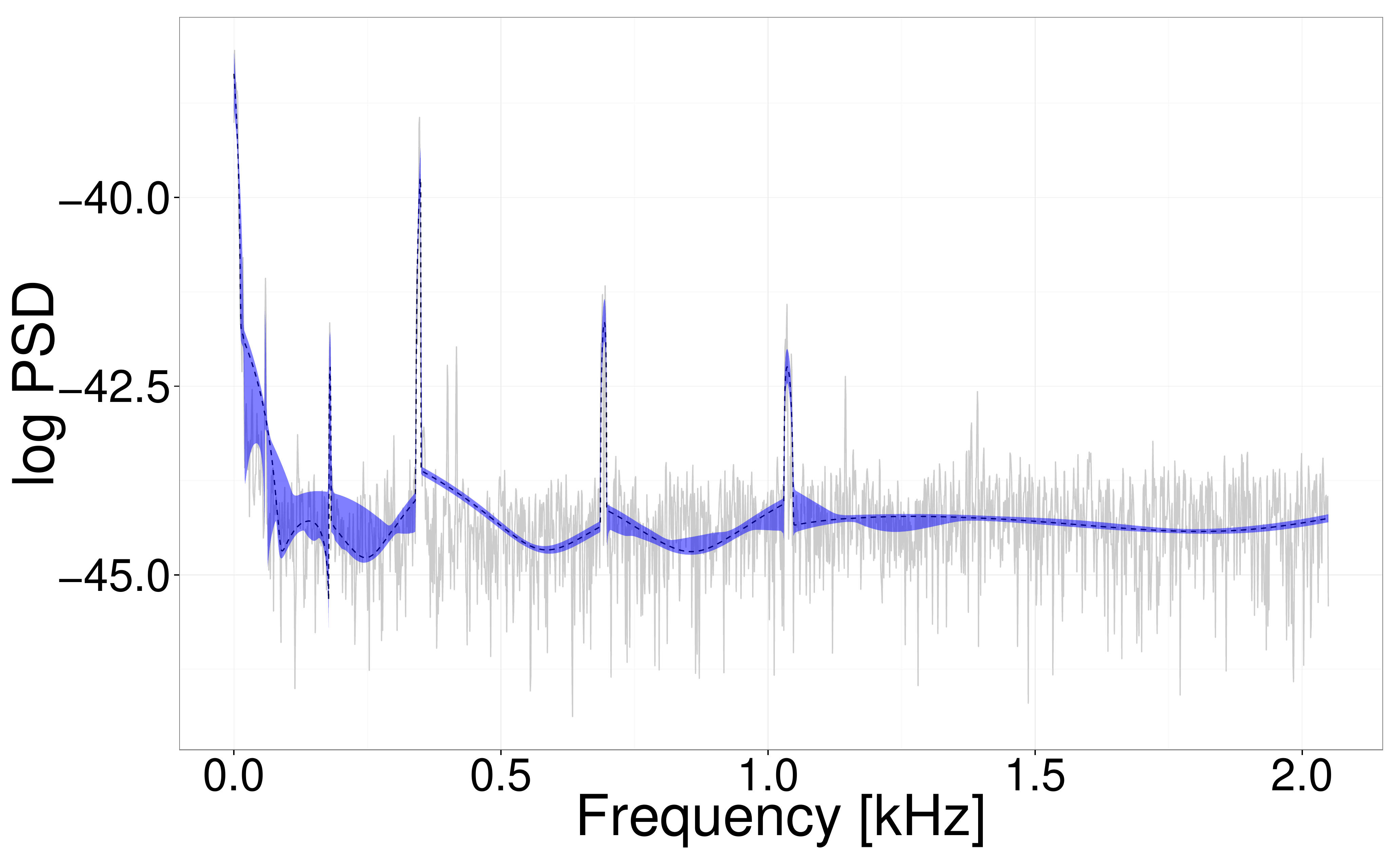}
\caption{Estimated $\mathrm{log}_{10}$-PSD for a 1~s segment of
  recoloured LIGO S6 data.  The posterior median log-PSD (dashed
  black) along with the 90\% pointwise credible region (shaded blue) are
  overlaid with the log-periodogram (grey).  The log transform is base
  10 here.}
\label{fig:psdligo}
\end{figure}

As demonstrated in Section~\ref{sec:simulation} (e.g.,
Figure~\ref{fig:ar4ci}), the Bernstein polynomial approach would have
struggled to estimate the abrupt changes of power present in real
detector data. It can be seen in Figure~\ref{fig:psdligo} though, that
the B-spline approach estimates the log-spectral density very well.
The estimated log-PSD follows the general broad-band shape of the
log-periodogram well, and the primary sharp changes in power are also
accurately estimated.  The method, however, seems to be less sensitive
to some of the smaller spikes.


\section{Conclusions and outlook}\label{sec:conclusions}

In this paper, we have presented a novel approach to spectral density
estimation, using a nonparametric B-spline prior with a variable
number and location of knots.  We have demonstrated that for
complicated PSDs, this method outperforms the Bernstein polynomial
prior in terms of IAE and uniform coverage probabilities.

The B-spline prior provides superior Monte Carlo estimates,
particularly for spectral densities with sharp and abrupt changes in
power.  This is not surprising as B-splines have local support and
better approximation properties than Bernstein polynomials.  However,
the favourable estimation qualities of the B-spline prior come at the
expense of increased computation time.

The posterior distribution of the B-spline mixture parameters with
variable number and location of knots could be sampled using the
RJMCMC algorithm of \citet{green:1995}, however RJMCMC methods are
often fraught with implementation difficulties, such as finding
efficient jump proposals when there are no natural choices for
trans-dimensional jumps \citep{brooks:2003}.  We avoid this
altogether by allowing for a data-driven choice of the smoothing
parameter and knot locations using the nonparametric Dirichlet process
prior.  This yields a much more straightforward sampling mechanism.

The B-spline prior was applied to the annual mean sunspot data set.  We
got a reasonable estimate of the log-PSD, and estimated that the solar
cycle occurs every 11.07 years.  This is consistent with existing
knowledge and previous analyses.

We have demonstrated that the B-spline prior provides a reasonable
estimate of the spectral density of real instrument noise from the
LIGO gravitational wave detectors.  In a future paper, we will focus
on characterizing this noise while simultaneously extracting a GW
signal, similar to \citet{edwards:2015}.  As the algorithm is
computationally expensive, it will be well-suited towards the shorter
burst-type signals (of order 1~s or less) like rotating core collapse
supernovae.  Using a large enough catalogue of waveforms,
estimation of astrophysically meaningful parameters could be done by
sampling from the posterior predictive distribution, similar to
\citet{edwards:2014}.  Another future initiative is to analyze the
impact of informative priors on the LIGO PSD estimates.

Though we have only presented the B-spline prior in terms of spectral
density estimation, it could be used in a much broader context, such
as in density estimation.  A paper using this approach for density
estimation is in preparation and could yield a more flexible,
alternative approach to the Triangular-Dirichlet prior function
\textsf{TDPdensity} in the \textsf{R} package \textsf{DPpackage}
\citep{jara:2011}.

\section*{Acknowledgements}

We thank Claudia Kirch, Alexander Meier, and Thomas Yee for fruitful
discussions, and Michael Coughlin for providing us with the recoloured
LIGO data.  We also thank the New Zealand eScience Infrastructure
(NeSI) for their high performance computing facilities, and the Centre
for eResearch at the University of Auckland for their technical
support.  NC's work is supported by National Science Foundation grant
PHY-1505373.  All analysis was conducted in \textsf{R}, an open-source
statistical software available on \textsf{CRAN} (cran.r-project.org).
We acknowledge the following \textsf{R} packages: \textsf{Rcpp, Rmpi,
  bsplinePsd, beyondWhittle, splines, signal, bspec, ggplot2, grid}
and \textsf{gridExtra}.  This paper carries LIGO document number
LIGO-P1600239.

\bibliographystyle{plainnat}
\bibliography{references}

\end{document}